\documentclass{article} %for one column style with line numbers
\usepackage{amsmath}

\usepackage{graphicx}

\begin{document}

\title{Transient Reviving Dynamics with an Exact Solution for Delay Differential Equations}

\author{Kenta Ohira$^{1}$ and Toru Ohira$^{2}$\\
\small \ $^{1}$Future Value Creation Research Center,\\ 
\small Graduate School of Informatics, Nagoya University, Japan\\
\small\ $^{2}$Graduate School of Mathematics, Nagoya University, Japan
}

\maketitle

\begin{abstract}
We present a new approach to examine transient dynamics in a class of non-autonomous delay differential equations. Exact solutions for these equations are obtained using the Lambert $W$ function alongside an appropriately chosen initial function. These solutions provide a reliable approximation for transient dynamics when the initial functions are not markedly distinct. We explore a non-autonomous equation that exhibits a distinctive phenomenon of reviving dynamics as an illustrative example. The derived exact solutions effectively encapsulate the qualitative characteristics of this reviving dynamics over various delay values.
\end{abstract}


%%% Keywords are not needed any longer. %%%
%%%\kword{keyword1, keyword2, keyword3, \ldots}
%%%

\section{Introduction}

Delays are inherent in various systems, particularly those involving feedback mechanisms. These delay elements are known to introduce complex behaviors into the system, leading to extensive research across disciplines such as mathematics, physics, biology, engineering, and economics \cite{heiden1979,bellman1963,cabrera1,hayes1950,insperger,kcuhler,longtinmilton1989a,mackeyglass1977,glass1988,glassmackey1988,miltonetal2009b,ohirayamane2000,smith2010,stepan1989,stepaninsperger,szydlowski2010}. A primary mathematical tool employed in these studies is delay differential equations. One well-known example is the Mackey--Glass delay differential equation \cite{mackeyglass1977}, initially proposed as a model for red blood cell regeneration, incorporating a delay in the regeneration process. A characteristic feature of the solution to this equation is the onset of oscillations as the delay increases, resulting in the destabilization of fixed points and the emergence of complex chaotic behavior (e.g.\cite{taylor}). While research in delay differential equations primarily focuses on stability conditions for solutions, many aspects, especially regarding transient behavior, remain largely unexplored\cite{milton_mmnp,pakdamanetal1998a,cantisan}.

Recently, we have started  investigating delay differential equations with time-varying coefficients and their transient bheviors \cite{kentaohira2022,kentaohira2023,kentaohira2023b}. This equation falls into a category known as non-autonomous delay differential equations (non-autonomous DDEs), which have been studied in a broader context. Once again, solving non-autonomous DDEs presents significant mathematical challenges\cite{Busenberg1984,Ming1990,Ford2002,Gyori2017}. In this context, we have identified, within our special case of non-autonomous DDEs, the potential occurrence of resonance phenomena in transient dynamical trajectories as a function of frequency or amplitude, with the delay acting as a parameter. Furthermore, by leveraging the Lambert $W$ function \cite{corless,shinozaki,pusenjak2017}, we have derived approximate solutions and demonstrated the ability to capture the characteristics of transient dynamical trajectories within a certain range of delay values.

Building upon this foundation, we extend our previous works to identify a class of non-autonomous DDEs for which exact solutions can be constructed using the $W$ function in conjunction with appropriately chosen initial conditions. This enables us to explore transient delayed dynamics across a range of DDEs. As an illustration, we propose a delay differential equation herein that exhibits a novel type of dynamical behavior. Particularly noteworthy is the transient amplitude ``reviving" dynamics observed within certain delay value ranges. In this phenomenon, the amplitude initially decreases towards a fixed point up to a specific time point but eventually diverges. To the best of the authors' knowledge, this transient amplitude reviving dynamics has not been previously identified or investigated within the context of delay differential equations. We conduct both analytical analysis employing the aforementioned exact solution construction with the $W$ function and numerical simulations to validate our approach. Finally, we discuss its relevance to resonant behaviors induced by delays.

\section{Non-autonomous delay differential equation}

%%%%%%%%

The delay differential equation we examine is of the form:
\begin{equation}
{dX(t)\over dt} + g(t) X(t) = b h(t)X(t-\tau).
\label{dr}
\end{equation}
Here $g(t),$ and $h(t)$ represent real--valued functions of $t$, and $b$ and $\tau \geq 0$ are real parameters.

Interpreting 
$\tau$ as a delay parameter, this equation describes the deleyed dynamics of the variable 
$X(t)$.  This equation falls within the category of non-autonomous DDEs.
When $g(t) = \alpha$ and $h(t)=1$ with a constant $\alpha$, it can also be viewed as a well-studied first-order delay differential equation with constant coefficients, $\alpha$ and $\beta$, commonly known as Hayes's equation \cite{hayes1950}:
\begin{equation}
{dX(t)\over dt} + \alpha X(t) = \beta X(t-\tau).
\label{hayes}
\end{equation}

Recently, our investigations explored cases where $g(t) = at$ with constant $a$ and $h(t) = 1$,  revealing transient dynamics with frequency resonances under specific parameter ranges \cite{kentaohira2022}. Additionally, for  $g(t) = at$ and $h(t) = \exp[-a \tau t]$, the equation exhibits transient oscillations with amplitude resonance. In this scenario, we derived an approximate analytical solution using the Lambert $W$ function\cite{corless,shinozaki,pusenjak2017}, capturing the transient dynamics and resonance conditions\cite{kentaohira2023}.

We extend these findings to the generalized form of Equation (\ref{dr}), determining conditions for which exact solutions can be constructed using the $W$ function with appropriately chosen initial conditions. 

%%%%%%%%

By establishing a suitable relationship between 
$g(t)$ and $h(t)$, we express the solution of Equation (\ref{dr})  using the exponential factor and the solution of the special case of Hayes's Equation (\ref{hayes}). Specifically, the following statement holds:
\vspace{2em}

The solution of the delay differential equation

\begin{equation}
{dX(t)\over dt} + g(t) X(t) = b h(t)X(t-\tau),\quad h(t) = \exp[-\int_{t-\tau}^t g(s) ds]
\label{drg}
\end{equation}

can be expressed as 
\begin{equation}
X(t) = \exp[-\int^t g(s) ds] \hat{X}(t),
\label{drgsol}
\end{equation}

where $\hat{X}(t)$ is the solution of
\begin{equation}
{d\hat{X}(t)\over dt}  = b\hat{X}(t-\tau).
\label{shayes}
\end{equation}
\vspace{2em}

\noindent
The above can be verified by direct substitution of (\ref{drgsol}) into  (\ref{drg}).

%%%%%%%%%%%

\subsection{Formal solution using the $W$ function}

Utilizing the property that the formal solution $\hat{X}$ of (\ref{shayes}) can be expressed using the Lambert $W$ function\cite{shinozaki,pusenjak2017}, which is defined as a multivalued complex function $W: C \rightarrow C$ satisfying
\begin{equation}
W(z)e^{W(z)} = z,
\label{wf}
\end{equation}
with branches denoted as $W_k, k=0, \pm 1, \pm 2, \dots, \pm \infty$, 
the general solutions $\hat{X}$ of (\ref{shayes}) can be written as follows:
\begin{equation}
\hat{X} = \sum_{k = -\infty}^{\infty} C_k e^{\lambda_k t}, \quad \lambda_k = {1 \over \tau} W_k (b \tau).
\label{solwf1}
\end{equation}
We observe that $\lambda_k$ are the roots of the transcendental characteristic equation given by:
\begin{equation}
\lambda  = b e^{-\tau\lambda}.
\label{chara}
\end{equation}
The constant coefficients $C_k$ are determined by the initial interval function $X(t) = \phi(t)$  over the interval $[-\tau, 0]$ of equation (\ref{shayes}).

Bringing everything together, the general solutions of the non-autonomous delay differential equation: 
\begin{equation}
{dX(t)\over dt} + g(t) X(t) = b \exp[-\int_{t-\tau}^t g(s) ds] X(t-\tau) 
\label{drg2}
\end{equation}
are formally expressed as:
\begin{equation}
X(t) = \exp[-\int^t g(s) ds] \sum_{k = -\infty}^{\infty} C_k e^{\lambda_k t}, \quad \lambda_k = {1 \over \tau} W_k (b \tau).
\label{soldrg2}
\end{equation} 

\subsection{Construction of exact solutions}

Typically, solving DDEs involves prescribing the initial function $X(t) = \phi(t)$  over the interval $[-\tau, 0]$.  Therefore, 
as we mention the coefficient $C_k$ needs to be calculated so that equation (\ref{soldrg2}) becomes a solution by requiring
\begin{equation}
 \phi(t) = \exp[-\int^t g(s) ds] \sum_{k = -\infty}^{\infty} C_k e^{\lambda_k t}, \quad t \in [-\tau, 0].
\label{solcond2}
\end{equation} 
Solving (\ref{solcond2}) for the coefficient $C_k$ analytically is challenging, often relying on numerical analysis\cite{pusenjak2017}.

However, in cases where specific dynamical behaviors or properties can be approximated by a single or a few terms out of the infinite sum in  (\ref{soldrg2}), we can work in reverse. In such instances, we choose the initial function $\phi(t)$ such that equation (\ref{solcond2}) holds for given $C_k$. Then for such initial function, equation (\ref{soldrg2}) serves as an exact solution. For example, if we utilize only 
the principal branch $W_0$, then
\begin{equation}
{X_{W_0}}(t) = {X_0}\exp[-\int^t g(s) ds] e^{{1 \over \tau} {W_0} (b \tau) t}
\label{soldrgw0}
\end{equation} 
with some constant $X_0$ is an exact solution of equation (\ref{drg2}) including the initial function for $t \in [-\tau, \infty]$.
Additionally, if the initial function can be approximated by this function for $t \in [-\tau, 0]$, it serves as an approximation for the solution of equation  (\ref{drg2})
for $0 < t$ to a certain degree. We demonstrate the efficacy of this approach with an example.

\subsection{Example}
An illustrative example from a previous study\cite{kentaohira2023} involves the case where $g(t) = at$. Then, the non-autonomous DDE (\ref{drg2}) becomes 
\begin{equation}
{dX(t)\over dt} + a t X(t) = b e^{- {1\over 2}a {\tau}^2} e^{-  a t \tau } X(t-\tau).
\label{drgeg}
\end{equation}
An exact solution for this equation for $t \in [-\tau, \infty]$ using the principle branch is given as follows:
\begin{equation}
{X_{W_0}}(t) = X_0 e^{- {1\over 2}a t^2} Re[e^{{1 \over \tau} W_0 (b \tau) t}] = X_0 e^{- {1\over 2}a t^2} e^{{1 \over \tau} Re[W_0 (b \tau)] t}
\cos({1 \over \tau} Im[W_0 (b \tau)] t).
\label{soldrgeg}
\end{equation}
This solution can also approximate the solution where the initial function $\phi(t)$ is not too different from above for $t \in [-\tau, 0]$.
For example, if we choose a constant initial function $\phi(t)=  X_0 , t \in [-\tau, 0]$, it can provide a rather good approximation for certain parameter ranges. An example is shown in Figure 1 with parematers $a=0.01, b= - 2.0$. We observe that equation (\ref{soldrgeg}) captures the transient dynamics quite well for $\tau <~ 2$.
\begin{figure}
\begin{center}
\includegraphics[height=12cm]{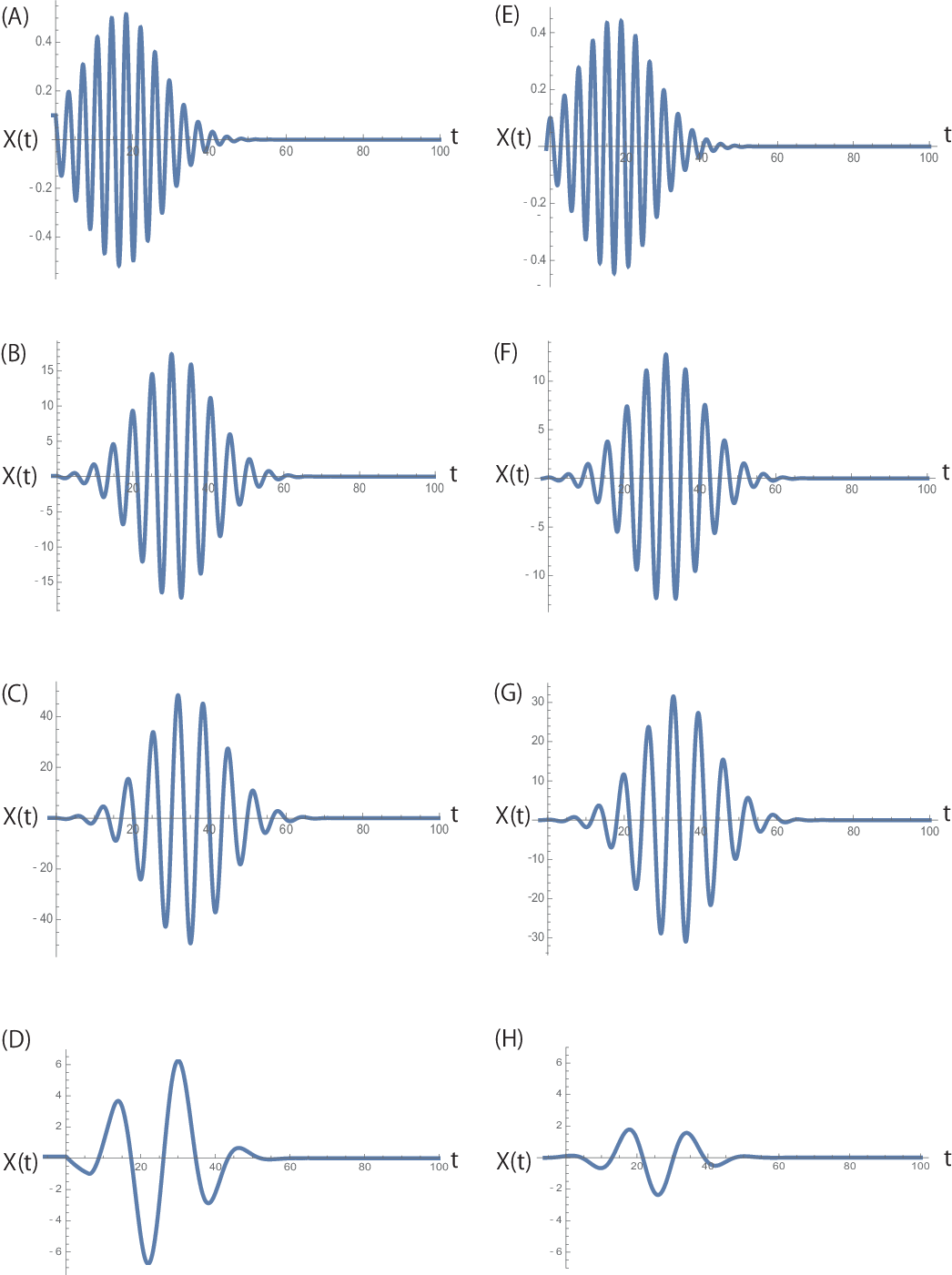}
\caption{Comparison of Dynamics of Equation (\ref{drgeg}) with a Constant Initial Function via Numerical Simulation (Left A-D) and Corresponding Dynamics given by Equation (\ref{soldrgeg}) (Right E-H):  The parameters are set as $a= 0.01, b= -2.0$ For (A-D), a constant initial interval function is set as $X(t) = 1.0$, $(-\tau \leq t \leq 0)$. The values of the delays $\tau$ are (A)(E) $1.0$, (B)(F) $1.5$, (C)(G) $2.0$, (D)(H) $6.0$.}
\label{dynamics}
\end{center}
\end{figure}

\section{Reviving Delayed Dynamics}

The procedure of constructing an exact solution as described above enables us to examine the behavior of transient dynamics for a class of non-autonomous DDEs. As an example, we introduce and examine a specific case of equation  (\ref{drg}) where $g(t) = a / (t +\tau_M)$:
\begin{equation}
{dX(t)\over dt} + {a}{({1 \over {t + \tau_M}})} X(t) = b (1- {\tau \over {t + \tau_M}})^a X(t-\tau).
\label{drs2}
\end{equation}
Here, $a$, $b$ and $\tau_M$ are real--valued constants, and we consider the range of delay $0 \leq \tau < \tau_M$. 
Our interest is to investigate the behavior of solution of this DDE for the range of $-\tau \leq t$. 

We employ our exact solution construction described in the previous section. The formal solution for Equation (\ref{drs2}) can be expressed using Equation (\ref{soldrg2}):
%%%%%%%%
\begin{equation}
X(t) = ({1 \over {t + \tau_M}})^{a} \sum_{k = -\infty}^{\infty} C_k e^{\lambda_k t}, \quad \lambda_k = {1 \over \tau} W_k (b \tau).
\label{soldrg2s}
\end{equation} 

Then, using  only the principal branch ($k = 0$) of the $W$ function, an exact solution for $-\tau \leq t$ is given as follows with a normalization such that it is $X_0$ at $t=0$:
\begin{equation}
{X_{W_0}}(t) = X_0 ({{\tau_M} \over {t+\tau_M}})^{a} Re[e^{{1 \over \tau} W_0 (b \tau) t}] = X_0 ({{\tau_M} \over {t+\tau_M}})^{a} e^{{1 \over \tau} Re[W_0 (b \tau)] t}\cos({1 \over \tau} Im[W_0 (b \tau)] t),
\label{soldrgegs2}
\end{equation}

The most noteworthy characteristic of the solution to this delay differential equation is the resurgence of oscillatory transient dynamics, as illustrated in Fig. 2, which shows a representative plot of the solution 
(\ref{soldrgegs2}). As a confirmation of our approach, we also numerically solved Equation (\ref{drs2}) with the initial function $\phi(t) = {X_{W_0}}(t)$ for $-\tau \leq t \leq 0$. The results match our exact solution plots (see Appendix).

\begin{figure}
\begin{center}
\includegraphics[height=12cm]{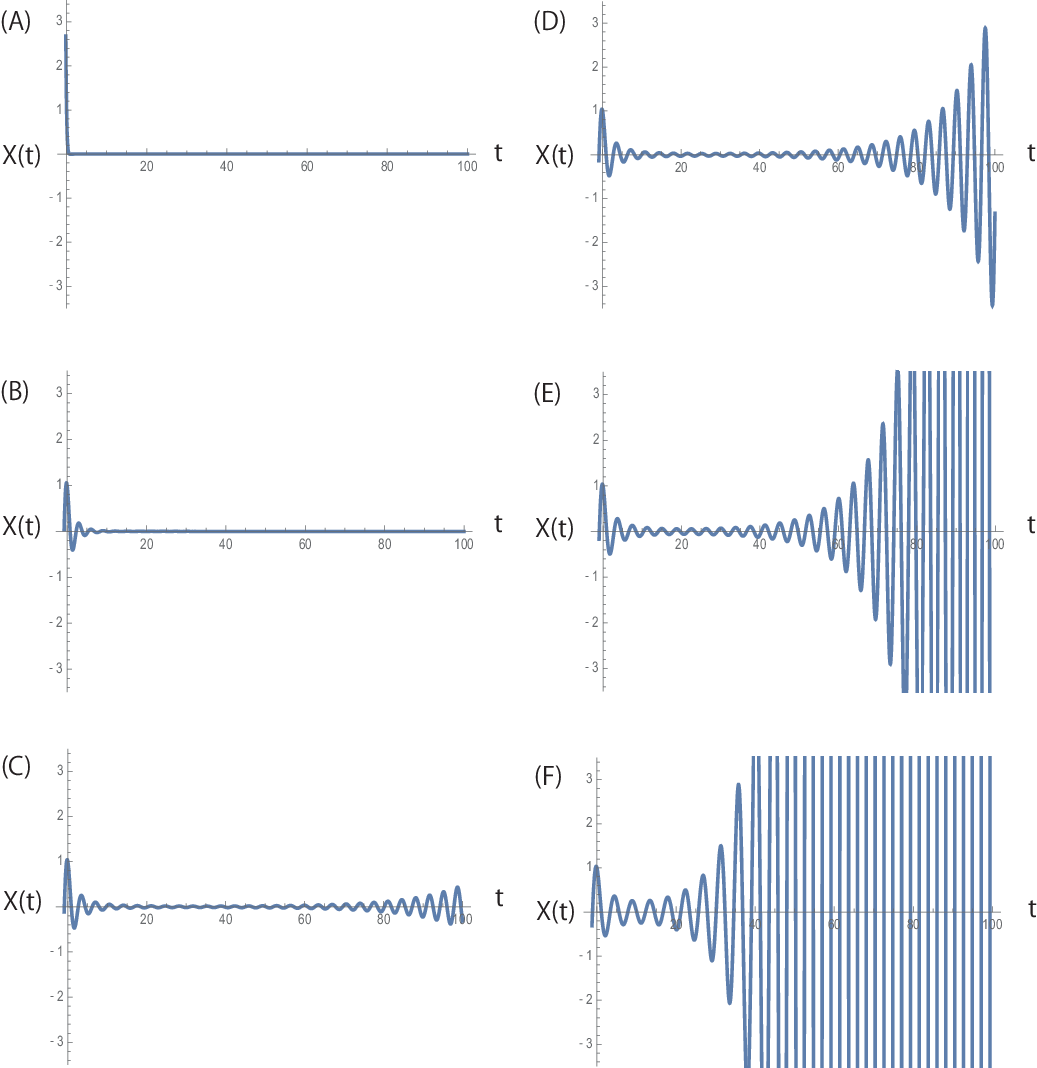}
\caption{Plots of the Solution (\ref{soldrgegs2}): The parameters are set as $a=5.0, b= -2.0, \tau_M = 8.0$. The values of the delays $\tau$ are (A)$0.2$, (B)$0.75$, (C)$0.92$, (D)$0.95$, (E)$1.0$, (F)$1.25$. (Part of trajectries are truncated in (E) and (F) beyond (-3.5, 3.5)).}
\label{dynamics}
\end{center}
\end{figure}

With appropriately chosen parameters, it displays a transient oscillation characterized by a diminishing amplitude followed by divergence. Unlike the typical behavior of oscillatory delayed dynamics generated by the first-order DDE, such as Hayes's equation (\ref{hayes}), where the amplitude monotonically decreases towards a fixed point or diverges beyond a critical delay value, the peculiar dynamics of the reviving amplitude have not been explicitly highlighted or explored.

%%%%%%%%%
From a qualitative standpoint, considering time 
$t$ in the equation's coefficients, the balance between dynamical increase and decrease evolves over time. By setting parameters 
$a$ and $b$ such that, for a given delay, the dynamics are asymptotically stable, the transient behavior decreases towards 
$X=0$. However, as time $t$ increases, the balance shifts, and the instability introduced by the delay becomes dominant, leading to divergence.

\section{Analysis of reviving dynamics}

In Equation (\ref{soldrgegs2}) , the amplitude of oscillation is determined by the function:
\begin{equation}
f(t) =  X_0 ({{\tau_M} \over {t+\tau_M}})^{a} e^{{1 \over \tau}Re[W_0 (b \tau)] t}.
\label{soldamp}
\end{equation}
To observe the change in this amplitude, we calculate its derivative:
\begin{equation}
{df(t) \over dt}=  - X_0 ({{\tau_M} \over {t+\tau_M}})^{a} ( {{a} \over {t+\tau_M}} - {1\over \tau} Re[W_0 (b \tau)] ) e^{{1 \over \tau}Re[W_0 (b \tau)] t} .
\label{soldamp}
\end{equation}
Here, we assume 
 $X_0 > 0$, $\tau \geq 0$ and consider $t > 0$ as in the parameters in Figure 2.
Then
\begin{equation}
{df(t) \over dt} < 0 \iff  Re[W_0 (b \tau)] < {{a\tau} \over {t+\tau_M}},
\label{leq0}
\end{equation}
and
\begin{equation}
{df(t) \over dt} = 0 \iff  Re[W_0 (b \tau)] = {{a\tau} \over {t+\tau_M}} .
\label{leq0}
\end{equation}
These lead to the following results:
\begin{equation}
{df(t) \over dt} < 0,  \quad \forall t >0 \iff  Re[W_0 (b \tau)] < 0.
\label{leqneg}
\end{equation}
The time $t_{min}$ that minimizes the amplitude is determined by
\begin{equation}
{df \over dt} (t_{min}) = 0.
\label{leqmint}
\end{equation}
Hence, 
\begin{equation}
t_{min} = { a \tau \over Re[W_0 (b \tau)] } - \tau_M.
\label{leqmint2}
\end{equation}

In Figure 3, we have plotted (A) $Re[W_0 (b \tau)]$ and (C) $t_{min}$ as functions of $\tau$. 
It is also known from the property of the $W$ function that
if $\tau_c = - {\pi \over {2 b}}$, then $Re[W_0 (b \tau_c)] =0$.
Additionally, the frequency component in the solution, $\omega_{W_0} = Im[W_0 (b \tau)]/ {\tau}$, is plotted in Figure 3(B).
Note that  $Im[W_0 (b \tau)]=0$ for $\tau \leq - {1 \over {b e}}$ (see Figure 3(B)). 
\vspace{1.2em}

\begin{figure}[ht]
\begin{center}
\includegraphics[height=9.5cm]{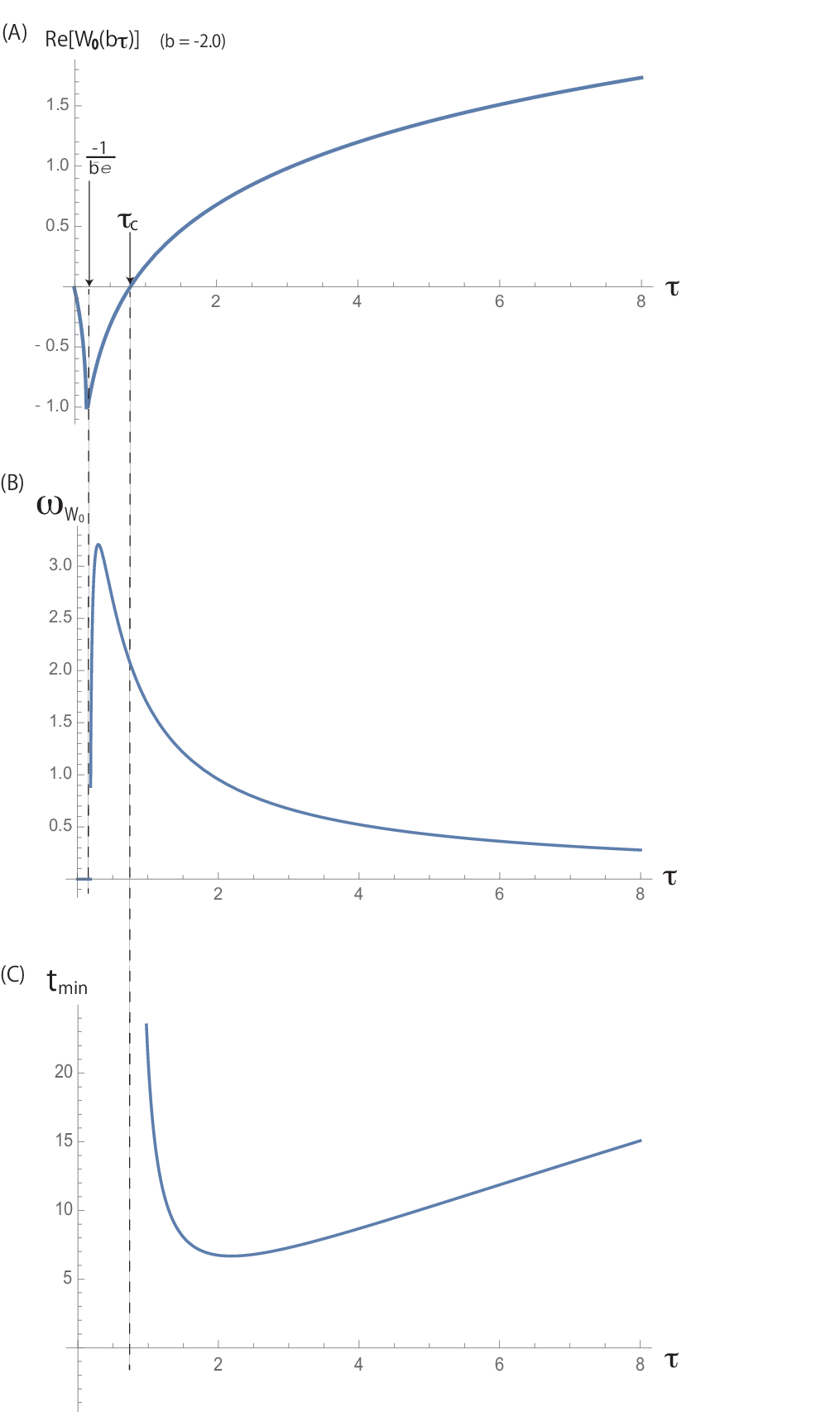}
\caption{Plots Relating to the Amplitude and Frequency of Solution (\ref{soldrgegs2}): The plots depict (A) the real part of $W_0$ function with $b= -2.0$, (B) the oscillation frequency $\omega_{W_0} = Im[W_0 (b \tau)]/ {\tau}$, and  (C) the time, $t_{min}$, at which the amplitude is at its minimum. 
We note $\omega_{W_0}=0$ for $\tau \leq {-1 \over {b e}}\approx 0.184$ and the maximum frequency $Max[\omega_{W_0}]=3.209$ is reached when
$\tau \approx 0.295$. 
The reviving dynamics occur for $\tau > \tau_c = - {\pi \over {2 b}} \approx 0.785$. The smallest value of $t_{min}$ is estimated to be $6.68$ when
$\tau \approx 2.182$.}
\label{wfunction}
\end{center}
\end{figure}

From these observations, we can infer the following:

\begin{itemize}

\item
For $0 \leq \tau \leq -{1 \over {b e}}$, $X=0$ is a stable fixed point, and the amplitude monotomically dicreases without oscillation.

\item
For $-{1 \over {b e}} < \tau < \tau_c$,$X=0$ is still a stable fixed point, but oscillaions appear.  The amplitude of oscillation monotonically decreases.

\item
The frequency, $\omega_{W_0}$,  increases with increasing delay, reaching a maximum frequency. Further increase in delay 
causes  $\omega_{W_0}$ to decrease.

\item
At $\tau \geq \tau_c$, the stability of $X=0$ is lost, and the dynamics diverge. However, reviving dynamics with minimum amplitude $X_{min}$ occur  at $t_{min}$. 
Som examples are indicated in Fig. 4.
%Some examples  are indicated in Fig. 3.

\item
For a range of parameters, there exists the smallest $t_{min}$ as a function of the delay $\tau$. In other words, tuning the value of the delay $\tau$ given $a, b, \tau_M$  allows for the shortest reviving switching point. 

\end{itemize}

\begin{figure}
\begin{center}
\includegraphics[height=8cm]{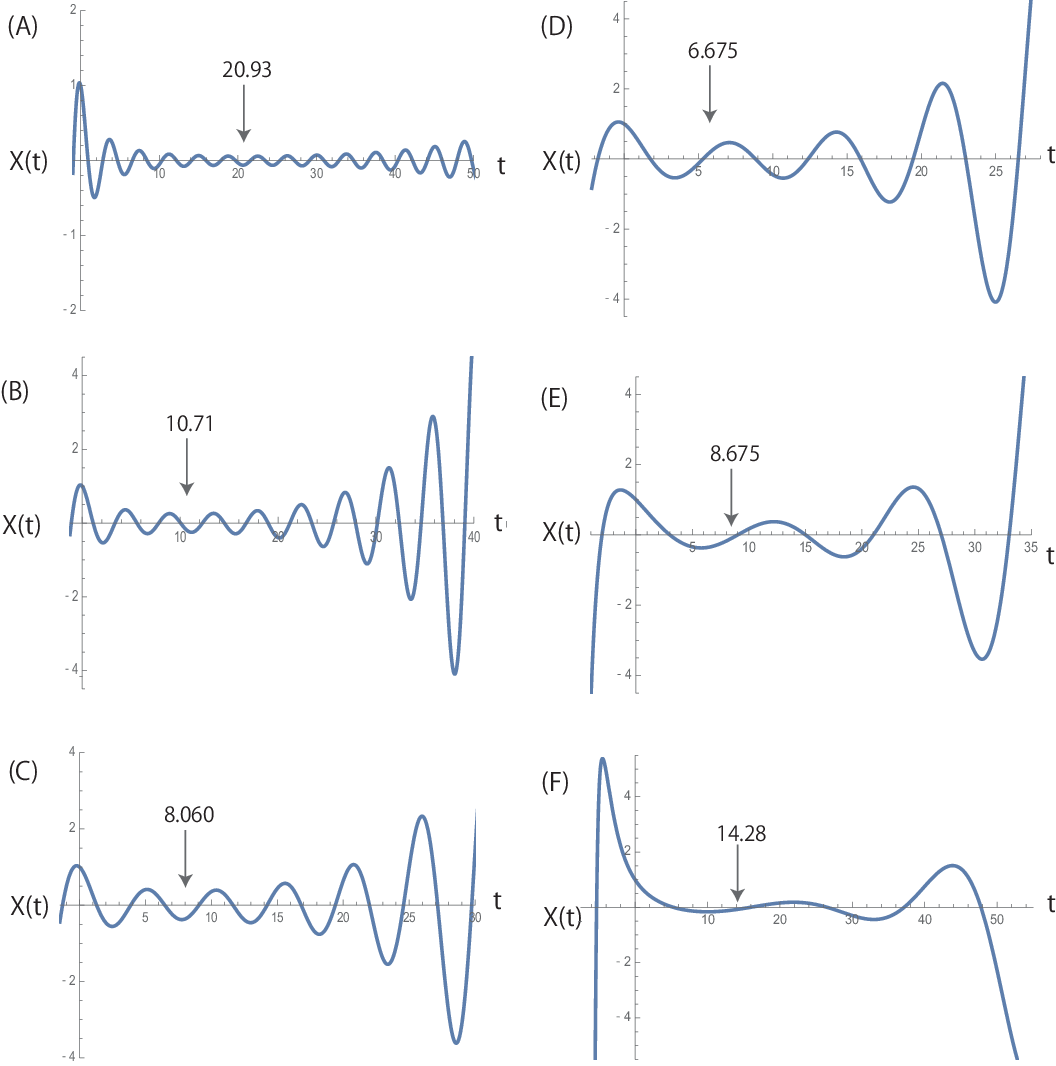}
\caption{Plots of Solution (\ref{soldrgegs2}) with the Minimum Amplitute Estimation: The parameters are set as $a=5.0, b= -2.0, \tau_M = 8.0$. The values of the delays $\tau$ are (A)$1.0$, (B)$1.25$, (C)$1.5$, (D)$2.18$, (E)$4$, (F)$7.5$.}
\label{wfunction}
\end{center}
\end{figure}

%%%%%%%%%%%%%%

\section{Discussion}

In this paper, we have expanded upon a formal solution using the $W$ function to encompass a class of non-autonomous DDEs. Additionally, we introduced a method to construct exact solutions for this class of DDEs and demonstrated its effectiveness in approximating dynamics when the initial function is not significantly different (see Appendix). 
In fact, even though we have only used the princeple branch of the $W$ function, one can use other branches or multiple combinations of branches if needed to approximate the pre-specified initial function. These advancements enable us to explore a wider range of peculiar behaviors arising from delays.

We discussed the phenomenon of reviving dynamics induced by delay as an exemplar of such intriguing behaviors. While previous studies have explored specific delay values inducing resonance in systems with delay feedback\cite{ohirasato1999,kentaohira2022,kentaohira2023}, the phenomenon examined in this paper also shares common traits, such as the emergence of a minimum value for the reviving time point.
We also note that oscillatory reviving dynamics can be obtained in other ways, such as simple harmonic oscillators with time-varying dissipative coefficients or other time-dependent potentials. Typically, however, these systems are described by second-order differential equations and/or with explicit oscillating external forces. As presented here, delays can induce this sort of reviving dynamics without such elements.

Our aim is that within the realm of delay differential equations, various transient phenomena can be further scrutinized. This analytical exploration may open up another pathway towards attaining a deeper understanding of the intricate dynamics within delay systems.

%%%%%%%%%%%%%%
\noindent
{\bf Acknowledgments}

The authors would like to thank Prof. Hideki Ohira and the members of his research group at Nagoya University for their useful discussions. This work was supported by the Yocho-gaku Project sponsored by Toyota Motor Corporation, JSPS Topic-Setting Program to Advance Cutting-Edge Humanities and Social Sciences Research Grant Number JPJS00122674991, JSPS KAKENHI Grant Number 19H01201, and the Research Institute for Mathematical Sciences, an International Joint Usage/Research Center located at Kyoto University.

\section*{Appendix}

In this appendix, we show results of direct numerical simulations of the delay differential equation  (\ref{drs2}) for the purpose of comparison against the exact solution given in Equation  (\ref{soldrgegs2}).

In Figure A-1, we plot the results with the same parameter settings as in Figure 2, with the matching initial function given by Equation (\ref{soldrgegs2}) for $-\tau \leq t \leq 0$. The results agree with those in Figure 2, confirming our approach.

We also plot, in Figure A-2, the results of the numerical simulation of Equation (\ref{drs2}) with the same parameter settings, except for an initial function which is taken as a constant $\phi(t) = 1.0$ for $-\tau \leq t \leq 0$. The discrepancies due to the difference in initial functions are shown in the enlarged example in Figure A-3.

As observed in the example in Section 2.3, the observed dynamical behavior of Equation (\ref{drs2}) for the constant initial function is qualitatively well-captured by Equation  (\ref{soldrgegs2}), including the reviving dynamics. Thus, there are cases where we can infer the qualitative transient behaviors of non-autonomous DDEs using our approach of utilizing a certain branch of the $W$ function.

\begin{figure}[h]
\begin{center}
\includegraphics[height=12cm]{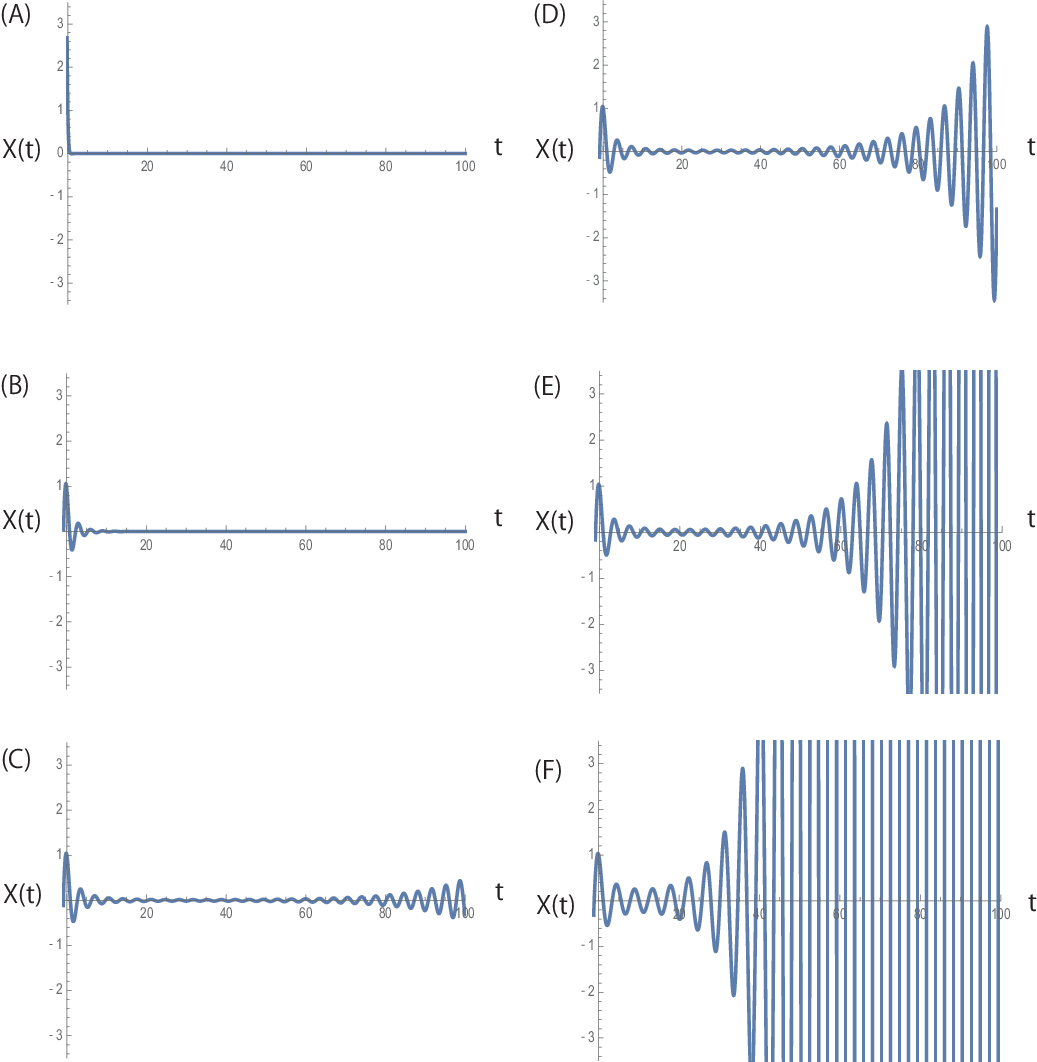}
\caption{Plots of the Solution of (\ref{drs2}) by Numerical Simulations with Inital Function Given by (\ref{soldrgegs2}): 
The parameters are consistent with those  in Figure 2, where $a=5.0, b= -2.0, \tau_M = 8.0$. The values of the delays $\tau$ are (A)$0.2$, (B)$0.75$, (C)$0.92$, (D)$0.95$, (E)$1.0$, (F)$1.25$. (Parts of trajectries are truncated in (E) and (F) beyond (-3.5, 3.5))}.
\label{a1}
\end{center}
\end{figure}

\begin{figure}
\begin{center}
\includegraphics[height=12cm]{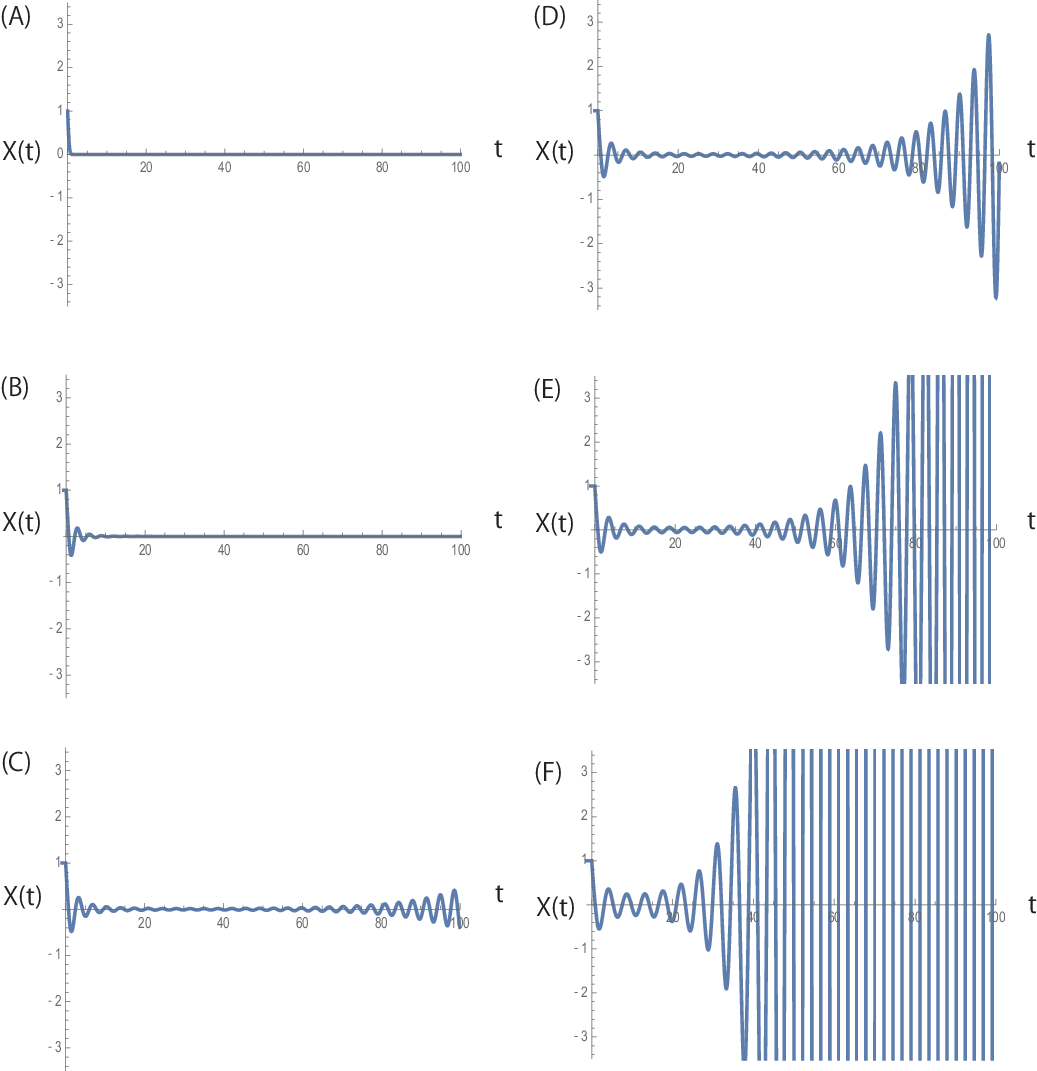}
\caption{Plots of the Solution of (\ref{drs2}) by Numerical Simulations with Constant Inital Function $\phi(t) = 1.0$: 
The parameters are consistent with those  in Figure 2, where $a=5.0, b= -2.0, \tau_M = 8.0$. The values of the delays $\tau$ are (A)$0.2$, (B)$0.75$, (C)$0.92$, (D)$0.95$, (E)$1.0$, (F)$1.25$. (Parts of trajectries are truncated in (E) and (F) beyond (-3.5, 3.5)).}
\label{a2}
\end{center}
\end{figure}

\begin{figure}
\begin{center}
\includegraphics[height=12cm]{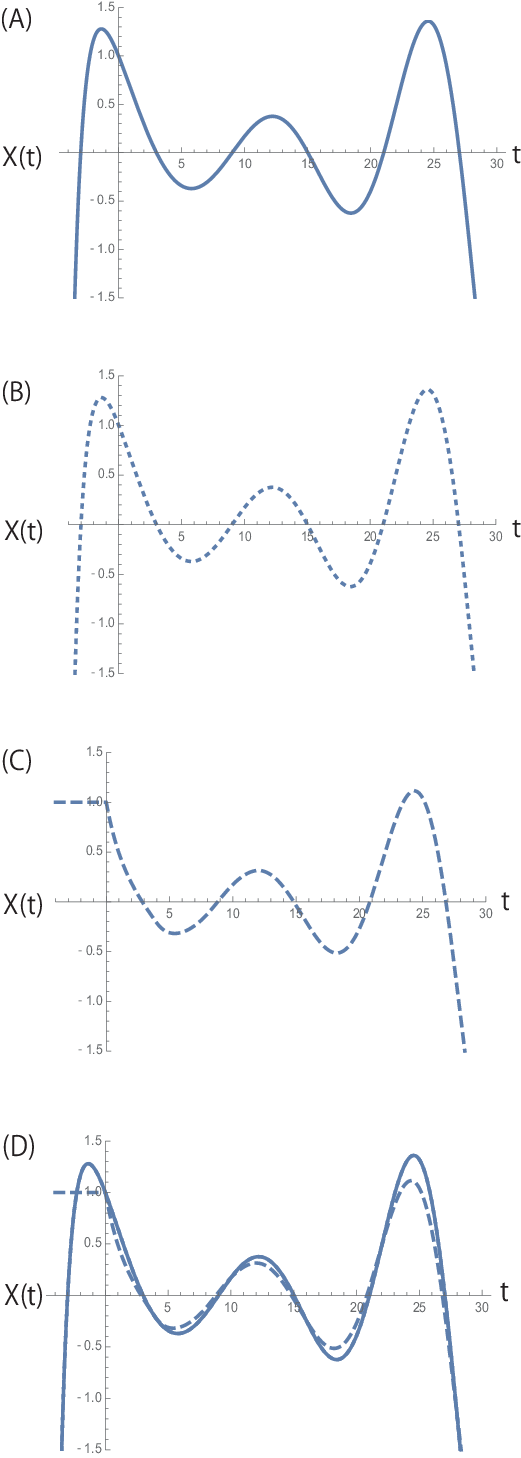}
\caption{Comparison of the Solutions of (\ref{drs2}): 
We compare the solutions obained through our analysis. The parameters are set as $a=5.0, b= -2.0, \tau_M = 8.0, \tau = 4.0$. (A) The plot of  (\ref{soldrgegs2}) (solid line), (B) Numerical simulations with inital function given by (\ref{soldrgegs2}) (dots), (C) Numerical simulation with constant inital function $\phi(t) = 1.0$ (dashed line), (D) Overlay plot of (A),(B),(C). The plots of (A) and (B) overlap, while (C) shows discrepancies.
}
\label{a3}
\end{center}
\end{figure}


\begin{thebibliography}{9}

\bibitem{heiden1979}
U.~an~der Heiden,
% Delays in physiological systems.
J. Math. Biol. {\bf 8}, 345 (1979).
 
\bibitem{bellman1963}
R.~Bellman and K.~Cook,
 {\em Differential--Difference Equations}
(Academic Press, New York, 1963).

\bibitem{cabrera1}
J.~L. Cabrera and J.~G. Milton,
%On--off intermittency in a human balancing task.
Phys. Rev. Lett. {\bf 89},158702 (2002).

\bibitem{hayes1950}
N.~D. Hayes.
%Roots of the transcendental equation associated with a certain difference--differential equation.
J. Lond. Math. Soc. {\bf 25}, 226 (1950).

\bibitem{insperger}
T.~Insperger,
%Act-and-wait concept for continuous-time control systems with feedback delay.
IEEE Trans. Control Sys. Technol. {\bf 14}, 974 ( 2006).


\bibitem{kcuhler}
U.~K{\" u}chler and B.~Mensch,
% Langevin's stochastic differential equation extended by a time-delayed term.
Stoch. Stoch. Rep. {\bf40}, 23 (1992).

\bibitem{longtinmilton1989a}
A.~Longtin and J.~G. Milton,
%Insight into the transfer function, gain and oscillation onset for the pupil light reflex using delay-differential equations.
Biol. Cybern. {\bf 61}, 51 (1989).

\bibitem{mackeyglass1977}
M.~C. Mackey and L.~Glass,
%Oscillation and chaos in physiological control systems.
Science {\bf197}, 287 (1977).


%%%%%%%%%

\bibitem{glass1988}
L.~Glass, A.~Beuter, and D. Larocque,
%Time delays, oscillations, and chaos in physiological control systems.
Mathematical Biosciences {\bf 90}, 111 (1988).

\bibitem{glassmackey1988}
L.~Glass and M.~C. Mackey,
 {\em From Clocks to Chaos: The rhythms of life}
 (Princeton University Press, Princeton, New Jersey, 1988).

%%%%%%%%%


\bibitem{miltonetal2009b}
J~Milton, J.~L. Cabrera, T.~Ohira, S.~Tajima, Y.~Tonosaki, C.~W. Eurich, and
  S.~A. Campbell,
%The time--delayed inverted pendulum: {I}mplications for human balance control.
Chaos {\bf 19}, 026110 (2009).

\bibitem{ohirayamane2000}
T.~Ohira and T.~Yamane,
% Delayed stochastic systems.
 Phys. Rev. E {\bf 61}, 1247 (2000).

\bibitem{smith2010}
H.~Smith,
 {\em An introduction to delay differential equations with
  applications to the life sciences}
(Springer, New York, 2010).

\bibitem{stepan1989}
G.~St{e}p{a}n,
 {\em Retarded dynamical systems: {S}tability and characteristic
  functions}
(Wiley \& Sons, New York, 1989).

\bibitem{stepaninsperger}
G.~St{e}pan and T.~Insperger,
%Stability of time-periodic and delayed systems: a route to act-and-wait control.
Ann. Rev. Control {\bf 30},159 (2006).

\bibitem{szydlowski2010}
M.~Szydlowski and A.~Krawiec,
%The Kaldor--Kalecki model of business cycle as a two-dimensional dynamical system. 
J. Nonlinear Math. Phys. {\bf 8}, 266 (2010).

\bibitem{taylor}
S. R. Taylor and S.~A. Campbell,
%Approximating chaotic saddles for delay differential equations.
Phys. Rev. E {\bf 75}, 046215 (2007).

%%%%%%%%
\bibitem{milton_mmnp}
J.~Milton, P.~Naik, C.~Chan, and S.~A. Campbell,
% Indecision in neural decision making models.
Math. Model. Nat. Phenom.  {\bf  5}, 125 (2010).

\bibitem{pakdamanetal1998a}
K.~Pakdaman, C.~Grotta-Ragazzo, and C.~P. Malta,
%Transient regime duration in continuous-time neural networks with delay.
Phys. Rev. E {\bf 58}, 3623 (1998).

\bibitem{cantisan}
J. Cantis\'{a}n, J, M. Seoane and M. A. F Sanju\'{a}n.
%``Transient chaos in time-delayed systems subjected to parameter drift,"
 {\em J.Phys.Complex.}, 2, 025001 (2021).

%%%%%%%
 
 \bibitem{kentaohira2022}
K.~Ohira.
Resonating Delay Equation.
 {\em EPL}, 137: 23001, 2022.
 
 \bibitem{kentaohira2023}
K.~Ohira and T.~Ohira.
Delayed Dynamics with Transient Resonating Oscillations.
 {\em J. Phys. Soc. Japan}, 92: 064002, 2023.
 
  \bibitem{kentaohira2023b}
K.~Ohira and T.~Ohira.
Delay, resonance and the Lambert W function.
ArXive: 2301.13448 (accepted and to appear in  {\em Springer Proceedings in Physics}, 2023).

%%%%%%%
\bibitem{Busenberg1984}
S. N.~Busenberg and K. L.~Cooke,
``Stability conditons for linear non-autonomous delay differential equations,"
{\em Quarterly of Applied Mathematics}, 10, 295--396, (1984).

\bibitem{Ming1990}
L. LI. Ming,
``Stability for linear non autonomous delay differential equations,"
{\em Math. Comput. Modelling}, 10, 67--74, (1990).


\bibitem{Ford2002}
N. J. Ford, S. M. Verduyn Lunel,
``Characterising small solutions in delay differential equations through numerical approximations,"
{\em Applied Mathematics and Computation}, 131, 253--270, (2002).

\bibitem{Gyori2017}
I. Gy\''{o}ri and L. Horv\'{a}th,
``Sharp estimation for the solution of delay differential and hallway type inequalities,"
{\em Discrete and Continuous dynamical systems}, 37, 3211-3242, (2017).

%%%%%%%

  \bibitem{corless}
R. M. Corless, G. H. Gonnet, D. E. G. Hare, D. J. Jeffrey, D. E. Knuth. 
On the Lambert W function
{\em Advances in Computational Mathematics}, 5: 329--359, 1996.

\bibitem{shinozaki}
H. Shinozaki and T. Mori. 
Robust stability analysis of linear time delay system by Lambert W function. 
{\em Automatica}, 42: 1791--1799, 2006.

\bibitem{pusenjak2017}
R. Pusenjak.
 Application of Lambert function in the control of production systems with delay. 
{\em Int. J. Eng. Sci}, 6: 28--38, 2017.

  
%%%%%%%%%%

\bibitem{ohirasato1999}
T.~Ohira and Y.~Sato.
Resonance with noise and delay.
 {\em Phys. Rev. Lett.}, 82:2811--2815, 1999.

\end{thebibliography}
\end{document}